\documentclass[journal,twoside,web]{ieeecolor}
\usepackage{tmi}
\usepackage{cite}
\usepackage{amsmath,amssymb,amsfonts}
\usepackage{algorithmic}
\usepackage{graphicx}
\usepackage{textcomp}
\usepackage{multirow}
\usepackage{booktabs}
\usepackage{hyperref}

\usepackage{xcolor}
\usepackage[textwidth=1mm, tickmarkheight=0pt]{todonotes}
\definecolor{grey}{RGB}{130,130,130}
\presetkeys{todonotes}{color=blue!20, textcolor=grey, size=\scriptsize}{}
\usepackage{marginnote}
\let\oldmarginnote\marginnote
\renewcommand{\marginnote}[1]{\oldmarginnote{\scriptsize\color{grey}#1}}

\def\BibTeX{{\rm B\kern-.05em{\sc i\kern-.025em b}\kern-.08em
    T\kern-.1667em\lower.7ex\hbox{E}\kern-.125emX}}
\begin{document}
\title{Self-supervised feature learning for cardiac Cine MR image reconstruction}
\author{Siying Xu, Marcel Früh, Kerstin Hammernik, Andreas Lingg, Jens Kübler, Patrick Krumm, Daniel Rueckert \IEEEmembership{(Fellow, IEEE)}, Sergios Gatidis, and Thomas Küstner \IEEEmembership{(Member, IEEE)}
\thanks{This work was supported by the Deutsche Forschungsgemeinschaft (DFG, German Research Foundation) under Germany’s Excellence Strategy – EXC 2064/1 – Project number 390727645.}
\thanks{Siying Xu, Marcel Früh, Sergios Gatidis and Thomas Küstner are with Medical Image and Data Analysis (MIDAS.lab), Department of Diagnostic and Interventional Radiology, University of Tuebingen, Tuebingen, Germany (e-mail: \{siying.xu; marcel.frueh; sergios.gatidis; thomas.kuestner\}@med.uni-tuebingen.de). }
\thanks{Kerstin Hammernik and Daniel Rueckert are with School of Computation, Information and Technology, Technical University of Munich, Germany (e-mail: \{k.hammernik; daniel.rueckert\}@tum.de). }
\thanks{Andreas Lingg, Jens Kübler and Patrick Krumm are with Department of Diagnostic and Interventional Radiology, University of Tuebingen, Germany (e-mail: \{andreas.lingg; jens.kuebler; patrick.krumm\}@med.uni-tuebingen.de). }
\thanks{Daniel Rueckert is also with Klinikum Rechts der Isar, Technical University of Munich, Munich,Germany and the Department of Computing, Imperial College London, London, United Kingdom.}
\thanks{Sergios Gatidis is also with Department of Radiology, Stanford University, Stanford, California, USA}}

\maketitle

\begin{abstract}
We propose a self-supervised feature learning assisted reconstruction (SSFL-Recon) framework for MRI reconstruction to address the limitation of existing supervised learning methods. Although recent deep learning-based methods have shown promising performance in MRI reconstruction, most require fully-sampled images for supervised learning, which is challenging in practice considering long acquisition times under respiratory or organ motion. Moreover, nearly all fully-sampled datasets are obtained from conventional reconstruction of mildly accelerated datasets, thus potentially biasing the achievable performance. The numerous undersampled datasets with different accelerations in clinical practice, hence, remain underutilized. To address these issues, we first train a self-supervised feature extractor on undersampled images to learn sampling-insensitive features. The pre-learned features are subsequently embedded in the self-supervised reconstruction network to assist in removing artifacts. Experiments were conducted retrospectively on an in-house 2D cardiac Cine dataset, including 91 cardiovascular patients and 38 healthy subjects. The results demonstrate that the proposed SSFL-Recon framework outperforms existing self-supervised MRI reconstruction methods and even exhibits comparable or better performance to supervised learning up to $16\times$ retrospective undersampling. The feature learning strategy can effectively extract global representations, which have proven beneficial in removing artifacts and increasing generalization ability during reconstruction.
\end{abstract}


\section{Introduction}
\label{sec:introduction}
\IEEEPARstart{M}{agnetic} resonance imaging (MRI) is widely used in clinical practice because of its excellent resolution for soft tissues, being non-invasive and radiation-free. However, the long acquisition time can cause patient discomfort, limit equipment capacity, and increase the risk of motion artifacts due to patient movement. To accelerate MR imaging, significant endeavors have been made over the past few decades.

Parallel imaging (PI) \cite{b1,b2,b3,b4} reduces the imaging time by utilizing multiple receiving coils distributed at different spatial locations. PI can be carried out in the k-space using GRAPPA \cite{b2} or in the image domain, such as SENSE \cite{b1}. To guarantee acceptable image quality, the acceleration rate is limited by the number of receiver coils and is typically set between $2\times$ to $4\times$ in clinical routines.

Compressed sensing (CS) MRI utilizes the concept that the image can be represented in a sparse transform domain. A random undersampling of the phase-encoding steps allows for the acceleration of the imaging, which in turn results in incoherent and noise-like aliasing artifacts in the image. Non-linear reconstruction algorithms can recover the aliasing-free image. Typical regularizations include wavelet transform \cite{b5}, Total Variation (TV) \cite{b6,b7}, and dictionary learning \cite{b8,b9,b10}. Moreover, imaging can be further accelerated by combining CS and PI to jointly exploit image sparsity and coil sensitivity information \cite{b11}. Despite improved imaging speed and efficiency, inappropriate hyperparameter selection may lead to over- or under-regularization, and the transform domain may not fully represent the sparse transformation.

Recently, the advancements in deep learning (DL) methods have significantly improved the reconstruction efficiency and the reconstructed image quality. K-space learning techniques, such as ALOHA \cite{b12} and RAKI \cite{b13}, enable non-linear estimation of missing k-space data from acquired k-space samples. Image enhancement networks \cite{b14,b15,b16} learn the mapping relationship between aliased inputs and the corresponding artifact-free images. Physics-based unrolled networks, such as the deep cascade network \cite{b17}, VN-Net \cite{b18}, and CINENet \cite{b19}, unroll the iterative optimization process into a deep network with intermittent data consistency (DC) layers, ensuring data fidelity to the acquired samples. Building upon the physics-based unrolled networks, hybrid learning networks such as KIKI-net \cite{b20} and MD-CNN \cite{b21} include additional neural networks for k-space learning, enabling information learning in multiple domains. Although these reconstruction methods have shown significant improvements compared to PI and CS, we observed two common limitations: (i) Fully-sampled data dependency: Most existing DL-based approaches are trained in a supervised manner, relying on abundant fully-sampled high-resolution images, which is challenging or sometimes impractical considering the prolonged acquisition times, or the presence of artifacts caused by breathing or other involuntary movements during long scan time. \cite{b44,b45} (ii) Domain sensitivity: Trained models are usually sensitive to domain shifts, such as when applied to different acceleration rates or subjects. A varying data distribution under changing undersamplings and physiological differences among individuals can lead to unstable performance and decreased generalization ability. \cite{b46,b47}

Self-supervised learning (SSL) methods for MR image reconstruction have been explored to avoid reliance on fully-sampled data. Yaman et al. proposed SSDU \cite{b22}, which splits the undersampled k-space data into two disjoint sets, one used in the DC step, while the other set is used to define the training loss. To augment training samples, they proposed multi-mask SSDU \cite{b23}, dividing each k-space into multiple k-space subset pairs. Millard et al. \cite{b48} analytically analyzed the performance of SSDU using the Noiser2Noise \cite{b49} framework. They explored the undersampling and re-undersampling strategies during training and inference, concluding that similar sampling characteristics are beneficial for training, whereas a single undersampling is preferred for inference. Zhou et al. proposed DDSS \cite{b24}, a fully self-supervised approach for non-cartesian MRI reconstruction, implemented by enforcing both k-space self-similarity and image-level appearance consistency. Noise2Recon \cite{b25} proposes a consistency training method that can use undersampled data by enforcing the reconstruction consistency between undersampled scans and their noise-augmented counterparts. Furthermore, energy-based and score-based functions were explored to perform generative modeling \cite{b41,b42,b43}, offering flexibility by not assuming a fixed measurement process. However, these generative methods rely on noise modeling, which may be sensitive to different sampling patterns. Therefore, although existing SSL methods represent significant progress, they still face challenges such as domain sensitivity and were primarily investigated for static image reconstruction, which may not effectively capture spatio-temporal features of motion-resolved data.

Self-supervised feature learning is a promising approach in SSL, where image representations are learned from the dataset and then applied to downstream tasks \cite{b26,b27,b28}. Different views of the same image are fed into a joint embedding architecture to learn the underlying feature representations invariant or insensitive to various views. The main challenge of this architecture is information collapse, in which the two branches ignore the input and produce identical and constant outputs.\cite{b29} Contrastive learning \cite{b30,b27,b31} and information maximization methods \cite{b29,b32} are two main families of approaches that differ in how they prevent information collapse. Contrastive methods explicitly maximize the agreement between embeddings of similar images while pushing away those of dissimilar samples. In information maximization methods such as Barlow Twins \cite{b32}, every pair of variables of the embedding vectors are decorrelated to prevent informational collapse without dissimilar samples. Self-supervised feature learning has demonstrated its feature extraction capabilities in various tasks, but its application to MR image reconstruction remains challenging. Unlike classification problems \cite{reviewSeg}, where feature learning primarily captures categorical similarities, MRI reconstruction is a regression problem that requires the extraction of meaningful structural representations. Some methods \cite{b33,b34} have attempted to introduce contrastive losses into reconstruction networks, but they primarily use learned features as a training regularization rather than integrating them into the reconstruction process. The critical challenge of designing a feature learning network that can effectively learn deep features which are beneficial for the reconstruction and a reconstruction network that leverages these learned features remains unexplored.

In this work, we propose a self-supervised feature learning assisted reconstruction (SSFL-Recon) framework for MR image reconstruction. To overcome the limitations of existing SSL reconstruction methods confined to static images, we adopted motion-resolved cardiac Cine data as the dataset for this study. Our proposed approach only requires undersampled data, eliminating the need for fully-sampled datasets. We aim to learn feature embeddings insensitive to sampling patterns, thereby enhancing generalization ability. Specifically, our framework comprises two steps: (i) Self-supervised feature learning, in which a feature extractor is trained using either contrastive learning or information maximization method. (ii) Self-supervised reconstruction, in which the pre-learned sampling-insensitive features are embedded to assist the reconstruction process. A novel reconstruction loss, which contains image consistency and cross k-space loss, is proposed.

The main contributions of our work can be summarized as follows: (i) A novel self-supervised feature learning assisted reconstruction framework: We propose a dedicated feature learning strategy tailored for MRI reconstruction, enabling the extraction of sampling-insensitive features solely from undersampled data. To effectively leverage these learned features in the reconstruction process, we integrate them into the reconstruction network, which enhances reconstruction stability across various sampling accelerations. (ii) A novel self-supervised reconstruction loss: We introduce a self-supervised loss consisting of image consistency and cross k-space loss, which aligns with the supervised loss but avoids the dependency on fully-sampled datasets. The proposed framework outperforms other SSL reconstruction methods and even achieves similar or superior performance compared to supervised reconstructions, which, however, depend on fully-sampled datasets.

\section{Method}
The proposed SSFL-Recon framework consists of two steps: (i) self-supervised feature learning and (ii) self-supervised image reconstruction. The objective of the first step is to extract sampling-insensitive features from the undersampled data so that the self-supervised reconstruction process in the second step can benefit from the assistance of pre-learned features and can obtain more robust reconstructions under various undersampling domain shifts.

\subsection{Self-supervised feature learning}
\begin{figure*}[!t]
\centerline{\includegraphics[width=\textwidth]{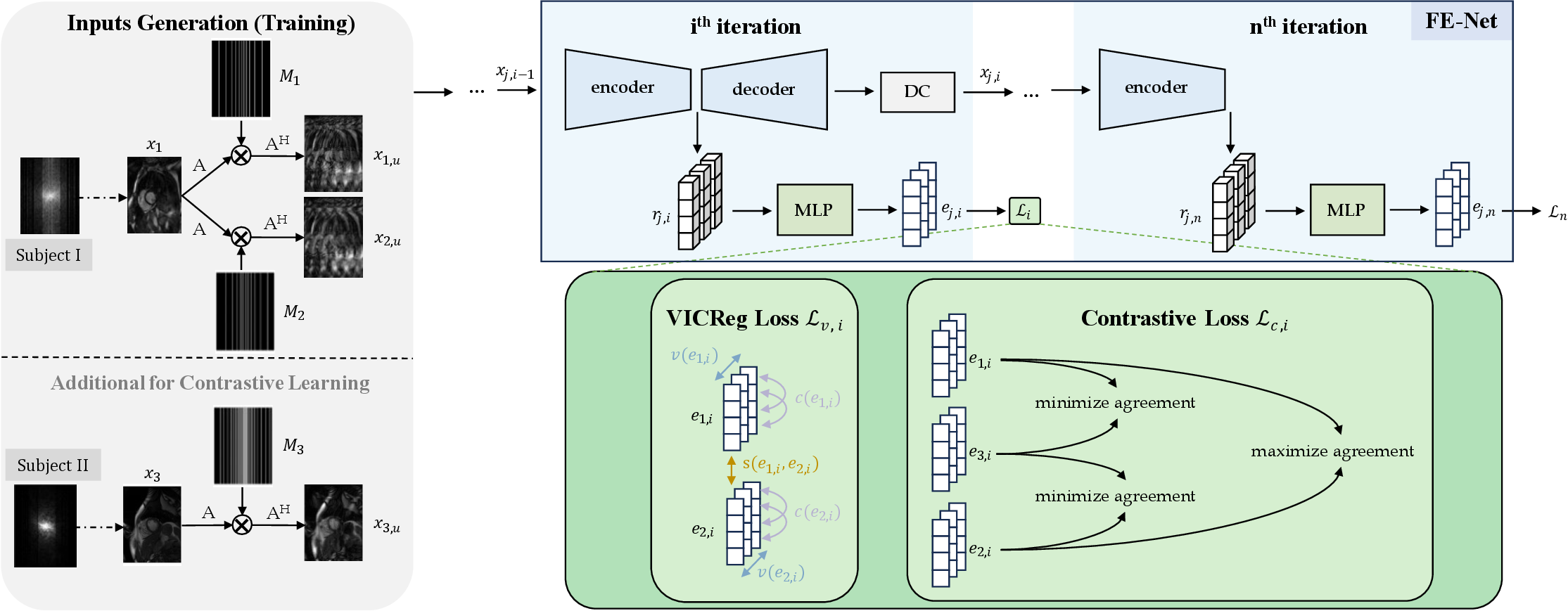}}
\caption{\textbf{The first step of the proposed SSFL-Recon framework: Self-supervised feature learning.} Input images $\mathbf{x}_{1,u}$ and $\mathbf{x}_{2,u}$ are generated from the same subject ($R=2$ conventional parallel imaging undersampled image $\mathbf{x}_{1}$) with two different sampling masks (contrastive learning: same acceleration rate but random generation seeds, information maximization method: random acceleration rate). The input image $\mathbf{x}_{3,u}$ is additionally needed for constructing negative pairs in contrastive learning, which is generated from another subject with an acceleration rate different from $\mathbf{M}_{1}$ and $\mathbf{M}_{2}$. The dashed arrows indicate the correspondence between the k-space and the image domain. The feature extraction network (FE-Net) is a physics-based unrolled network with $n$ iterations. In $i^{th}$ iteration, the encoder extracts the feature representation $\mathbf{r}_{j,i}$ from the input image $\mathbf{x}_{j,i-1}$, $j\in \left \{ 1,2,3 \right \} $. The multi-layer perceptron (MLP) projects $\mathbf{r}_{j,i}$ into a latent embedding space $\mathbf{e}_{j,i}$ to calculate feature learning loss $\mathcal{L}_{i}$, which can be VICReg loss $\mathcal{L}_{v, i}$ or contrastive loss $\mathcal{L}_{c, i}$. The VICReg regularization loss contains variance $v$, invariance $s$, and covariance $c$. The contrastive loss maximizes the agreement between positive pairs while minimizing the agreement between negative pairs.}
\label{fig:feature learning}
\end{figure*}

Similar to the concepts in computer vision \cite{b26,b27,b28,b29,b30,b31,b32}, we hypothesize that the extracted features from two different views of the same image, which are under different influences of aliasing artifacts in our study, share a common latent feature representation that is invariant or insensitive to the perspective, i.e., the embedded features are close to each other in the feature space. This section investigates two feature learning methods (contrastive learning and information maximization methods) to learn subject-specific and sampling-insensitive features from undersampled MR images. Fig.~\ref{fig:feature learning} illustrates the proposed feature learning step. To simplify the expression, we use SSFL-Recon(c) and SSFL-Recon(v) to distinguish between the two feature learning methods ('c' for contrastive learning, 'v' for the information maximization approach using VICReg loss \cite{b29}).

\subsubsection{Contrastive learning}
In contrastive learning, the positive pair is obtained from undersampled images of the same subject with similar undersampling characteristics. As shown in the inputs generation part in Fig.~\ref{fig:feature learning}, we construct positive pairs, $\mathbf{x}_{1,u}$ and $\mathbf{x}_{2,u}$, by re-undersampling $2\times$ conventional parallel imaging undersampled image $\mathbf{x}_{1}$, meaning that only a single prospective undersampled Cine dataset is required to create a positive pair. Specifically, we apply two random sampling masks, $\mathbf{M}_{1}$ and $\mathbf{M}_{2}$, which share the same acceleration rate but differ in generation seeds, to the complex-valued and coil-combined data to generate two re-undersampled k-spaces: 
\begin{equation}
    \mathbf{y}_{1}=\mathbf{A}_{1}\mathbf{x}_{1},\,\
    \mathbf{y}_{2}=\mathbf{A}_{2}\mathbf{x}_{1},
\label{eq:ForwardOp}
\end{equation}
in which $\mathbf{A}_{k}=\mathbf{M}_{k}\mathbf{FS}$ is the multi-coil forward encoding operator consisting of coil sensitivity maps $\mathbf{S}$, Fourier transformation $\mathbf{F}$, and sampling mask $\mathbf{M}_{k}$, $k\in \left \{ 1,2 \right \} $. Coil sensitivities are estimated from the low-resolution k-space center using ESPiRIT \cite{b4}. The adjoint operations are then applied to obtain two undersampled images, $\mathbf{x}_{1,u}$ and $\mathbf{x}_{2,u}$. These two images are from the same subject, referred to as positive pairs, i.e., two different views of the same image:
\begin{equation}
    \mathbf{x}_{1,u}=\mathbf{A}_{1}^\mathrm{H}\mathbf{y}_{1},\,\
    \mathbf{x}_{2,u}=\mathbf{A}_{2}^\mathrm{H}\mathbf{y}_{2}.
\label{eq:BackwardOp}
\end{equation}
The common and similar feature embeddings inherent in these two undersampled images represent the latent subject-specific and sampling-insensitive features.

Negative pairs are images with lower similarity, which are retrospectively obtained from different subjects with distinct undersampling patterns. In a prospective study, negative pairs can be acquired by utilizing acquisitions from different subjects with different samplings. Specifically, we obtain the undersampled image $\mathbf{x}_{3,u}$ from the image $\mathbf{x}_{3}$, which comes from a different subject than $\mathbf{x}_{1}$:
\begin{equation}
    \mathbf{x}_{3,u}=\mathbf{A}_{3}^\mathrm{H}\mathbf{A}_{3}\mathbf{x}_{3}.
\end{equation}
Consequently, $\mathbf{x}_{1,u}$ and $\mathbf{x}_{3,u}$, $\mathbf{x}_{2,u}$ and $\mathbf{x}_{3,u}$ form two negative pairs. To enhance their dissimilarity, the acceleration difference between sampling masks $\mathbf{M}_{1}$ and $\mathbf{M}_{3}$, as well as $\mathbf{M}_{2}$ and $\mathbf{M}_{3}$, is enforced to be larger than five.

In $i^{th}$ iteration of the feature extraction network (FE-Net), the encoder extracts the spatio-temporal feature representation $\mathbf{r}_{j,i} \in \mathbb{C}^{b\times t\times x\times y\times n_{f,e}}$ from the input image $\mathbf{x}_{j,i-1}$, $j\in \left \{ 1,2,3 \right \} $. Here, $b$ is the batch size, $(t, x, y)$ are the temporal and spatial sizes of the encoded feature, and $n_{f,e}$ is the feature/channel size after the last pooling layer in the encoder. Note that the feature learning loss is calculated for 2D spatial features. We treat the temporal dimension as independent samples in the batch dimension. The 2-channel real-valued numbers (concatenated real and imaginary components along the channel dimension of complex-valued representation) are projected by a multi-layer perceptron (MLP) into a latent embedding space $\mathbf{e}_{j,i}\in \mathbb{R}^{t\times n_{f,m}}$ for loss calculation, with $n_{f,m}$ being the feature/channel size after the MLP. For simplicity, we omitted the subscript $i$ in the subsequent loss functions. However, it is important to note that the loss is independently calculated for each iteration. The contrastive loss for $i^{th}$ iteration follows Information Noise Contrastive Estimation (InfoNCE) loss \cite{b35}:
\begin{equation}
    \mathcal{L}_{c,i} =-\frac{1}{N} {\textstyle \sum_{n=1}^{N}}  \mathrm{log} \frac{\mathrm{exp}(\mathrm {sim}(\mathbf{e}_1,\mathbf{e}_2)/\tau   ) }{\sum_{j=1}^{2}\sum_{m=j+1}^{3} \mathrm {sim}(\mathbf{e}_j,\mathbf{e}_m)/\tau }, 
\label{eq:contraloss}
\end{equation}
where $N$ is the number of samples in the training dataset, $\mathbf{e}_{j}$ is the feature embedding of the input image $\mathbf{x}_{j,u}$, $\mathrm{sim}(\mathbf{e}_{j},\mathbf{e}_{m})$ is the cosine similarity between vectors $\mathbf{e}_{j}$ and $\mathbf{e}_{m}$, and $\tau$ is a temperature coefficient used to smooth the exponential term.

\subsubsection{Information maximization method}
Besides contrastive learning, we explored information maximization methods, which dispense the need for negative pairs and solely necessitate similar instances. Here, we further broaden the sampling insensitivity of learned features by allowing $\mathbf{M}_{1}$ and $\mathbf{M}_{2}$ to have random acceleration rates. FE-Net remains the same as contrastive learning,  depicted in Fig.~\ref{fig:feature learning}.

The loss calculation follows variance-invariance-covariance regularization (VICReg) \cite{b29}, which is calculated at the embedding level on $\mathbf{e}_{1}=\left [ e_{1}^{1},\cdots ,e_{1}^{t} \right ]$ and $\mathbf{e}_{2}=\left [ e_{2}^{1},\cdots ,e_{2}^{t} \right ]$ composed of $t$ vectors with dimension $c$:
\begin{equation}
\begin{split}
    \mathcal{L}_{v,i} = \lambda s \left ( \mathbf{e}_{1}, \mathbf{e}_{2} \right ) +  \mu \left [ v\left ( \mathbf{e}_{1} \right ) + v\left ( \mathbf{e}_{2} \right ) \right ] + \nu \left [ c\left ( \mathbf{e}_{1} \right ) + c\left ( \mathbf{e}_{2} \right ) \right ],  
\label{eq:VICReg}
\end{split}
\end{equation}
where $s, v, c$ are the invariance, variance, and covariance regularization terms with weighting factors $\lambda, \mu, v$ controlling importance. Specifically, the invariance regularization term is defined as:
\begin{equation}
    s \left ( \mathbf{e}_{1}, \mathbf{e}_{2} \right )=\frac{1}{t} \sum_{j=1}^{t}\left \| e_{1}^{j} -e_{2}^{j}  \right \|_{2}^{2},   
\label{eq:invariance}
\end{equation}
which is the mean squared Euclidean distance between two vectors in each dimension. It encourages two vectors to be close, thus extracting similar features from two different views, i.e., similar representations under different aliasing artifacts.

The variance regularization term is defined as:
\begin{equation}
    v\left ( \mathbf{e}_{1} \right )=\frac{1}{c}\sum_{k=1}^{c}\max \left ( 0, \gamma -S\left ( \mathbf{e}_{1,k}, \epsilon  \right )  \right ),      
\label{eq:variance}
\end{equation}
where $\mathbf{e}_{1,k}$ is the vector composed of each value at dimension $k$ in all vectors in $\mathbf{e}_{1}$. $\gamma$ is the target constant value of the standard deviation, which is set to 1, following the setting in \cite{b29}, and $S$ is the regularized standard deviation:
\begin{equation}
    S\left ( \mathbf{e}_{1,k},\epsilon  \right ) =\left ( \mathrm{Var} \left (\mathbf{e}_{1,k} \right )+\epsilon \mathbf{I} \right ) ^{\frac{1}{2}} ,   
\label{eq:std}
\end{equation}
where $\epsilon$ is a small number used to prevent numerical instabilities \cite{b29}, and $\mathbf{I}$ is the identity matrix. This term enforces the variance along each dimension in one batch to be close to $\gamma$, which is necessary to prevent information collapse, meaning that all inputs are mapped to the same vector without valuable information, such as all zero.

The covariance regularization term is defined as:
\begin{equation}
    c\left ( \mathbf{e}_{1} \right )=\frac{1}{c}\sum_{m\ne n}^{}\left [ \mathbf{C}\left ( \mathbf{e}_{1}  \right )  \right ]_{m,n}^{2},       
\label{eq:covariance reg}
\end{equation}
with covariance matrix:
\begin{equation}
    \mathbf{C}\left ( \mathbf{e}_{1} \right )=\frac{1}{t-1}\sum_{j=1}^{t}\left ( e_{1}^{j} -\bar{e_{1}}   \right ) \left ( e_{1}^{j} -\bar{e_{1}}   \right )^\mathrm{T}, 
\label{eq:covariance}
\end{equation}
where $\bar{e}_{1}=\frac{1}{t}\sum_{j=1}^{t}e_{1}^{j}$. By encouraging the sum of non-diagonal elements in the covariance matrix to be close to 0, the features can be decorrelated, thus ensuring richer features.

\subsubsection{Network architecture}
As shown in Fig.~\ref{fig:feature learning}, FE-Net deploys an unrolled architecture to learn $n$ feature embeddings for $n$ iterations, which are utilized in the self-supervised reconstruction training afterwards. The same unrolled network architecture in both steps ensures that the extracted features are consistent with the noise level during the reconstruction. Each iteration contains one UNet comprised of an encoder and a decoder, one MLP, and one data consistency layer. The architectural details are the same as those of the reconstruction network, which will be introduced in the next section. We calculate the feature learning loss for each iteration, with the total loss being the sum of all $n$ iterations:
\begin{equation}
    \mathcal{L}=\mathcal{L}_{1}+\mathcal{L}_{2}+\cdots +\mathcal{L}_{n},
\label{eq:TotalLoss}
\end{equation}
where $\mathcal{L}_{i}$ is the feature learning loss (contrastive or VICReg loss) of the $i^{th}$ iteration. After pre-training, all MLPs are discarded, and the learned features extracted by each encoder are used for the reconstruction task.

\subsection{Self-supervised reconstruction}
\begin{figure*}[!t]
\centerline{\includegraphics[width=\textwidth]{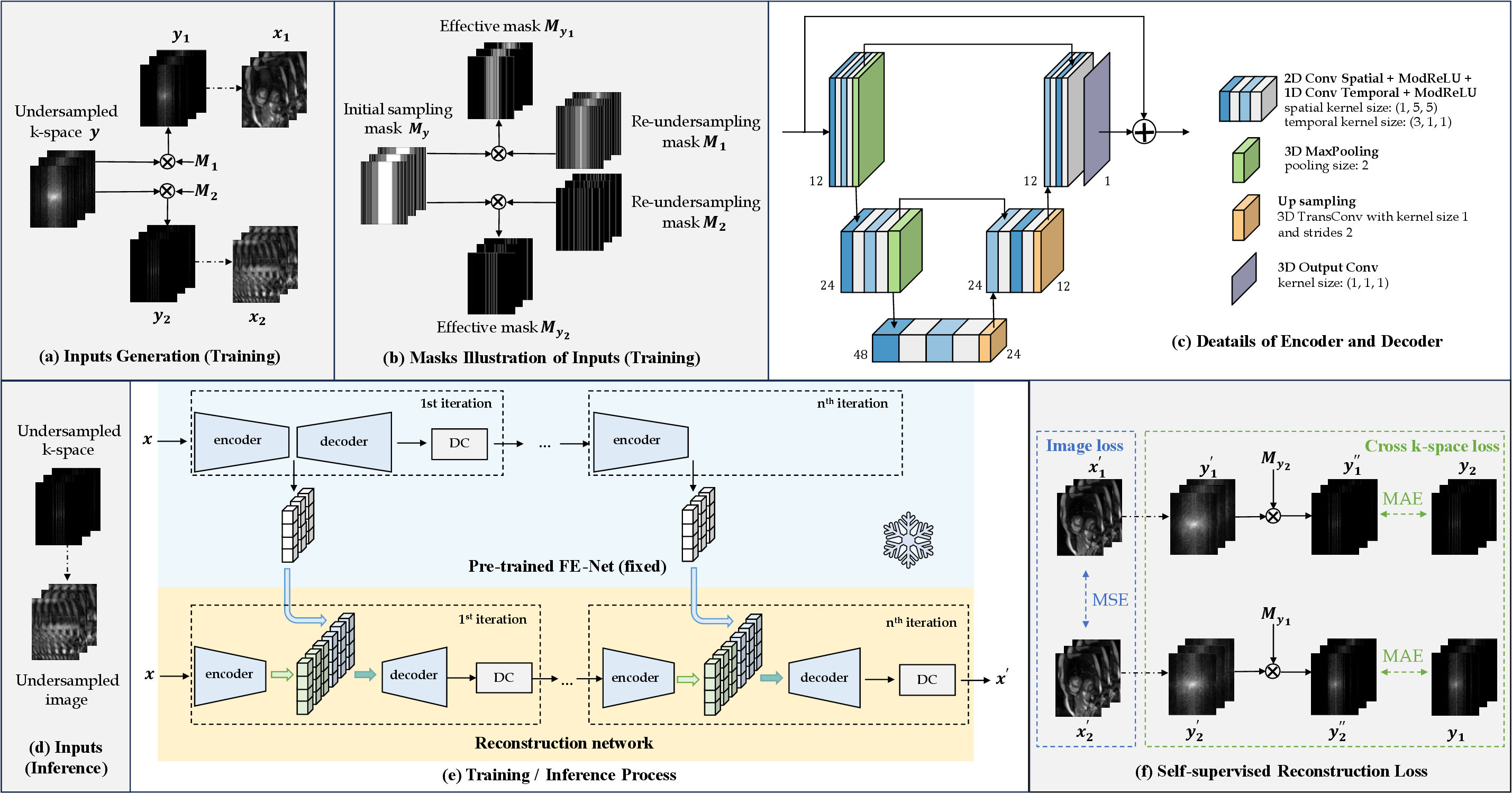}}
\caption{\textbf{The second step of the proposed SSFL-Recon framework: Self-supervised reconstruction.} (a) Inputs generation (training): two undersampled images, $\mathbf{x}_{1}$ and $\mathbf{x}_{2}$, are obtained by re-undersampling the accelerated k-space $\mathbf{y}$. Dashed arrows indicate the correspondence between the k-space and the image domain. (b) Masks illustration of training inputs: the initially undersampled k-space $\mathbf{y}$ corresponds to the undersampling mask $\mathbf{M}_{y}$. Re-undersampling masks $\mathbf{M}_{1}$ and $\mathbf{M}_{2}$ generate further undersampled k-spaces $\mathbf{y}_{1}$ and $\mathbf{y}_{2}$. The effective masks for them are denoted as $\mathbf{M}_{y_{1}}$ and $\mathbf{M}_{y_{2}}$. (c) Details of the encoder and decoder: the number of convolution kernels is annotated beside each layer. (d) Inputs for inference: singly undersampled data. (e) Training/inference process: The reconstruction network is a physics-based unrolled network with the same architecture as the feature learning step. In each iteration, the learned features from the pre-trained FE-Net (blue cubes) are concatenated with features of the reconstruction network (green cubes). Undersampled data $\mathbf{x}=\mathbf{x}_{1}/\mathbf{x}_{2}$ (Training) / $\mathbf{x}$ (Inference) are reconstructed to images $\mathbf{x}^{\prime}=\mathbf{x}_{1}^{\prime}/\mathbf{x}_{2}^{\prime}$ (Training) / $\mathbf{x}^{\prime}$ (Inference). (f) Self-supervised reconstruction loss: consisting of: 1) Image loss: mean squared error (MSE) between reconstructed images $\mathbf{x}_{1}^{\prime}$ and $\mathbf{x}_{2}^{\prime}$. 2) Cross k-space loss: mean absolute error (MAE) between reconstructed and sampled k-space points ($\mathbf{y}_{1}^{\prime\prime}$ and $\mathbf{y}_{2}$, $\mathbf{y}_{2}^{\prime\prime}$ and $\mathbf{y}_{1}$).} 
\label{fig:Reconstruction_Network}
\end{figure*}

The second step of the proposed SSFL-Recon framework is self-supervised reconstruction, which is illustrated in Fig.~\ref{fig:Reconstruction_Network} with the proposed image and k-space loss functions.

\subsubsection{Data pipeline and loss function}
As shown in Fig.~\ref{fig:Reconstruction_Network}(a), the accelerated k-space $\mathbf{y}$, which was undersampled with a mask $\mathbf{M}_{y}$, is re-undersampled to acquire two subsets of sampled points: $\mathbf{y}_{1}$ and $\mathbf{y}_{2}$. The re-undersampling masks are two independent masks, $\mathbf{M}_{1}$ and $\mathbf{M}_{2}$, with random acceleration rates shown in Fig.~\ref{fig:Reconstruction_Network}(b). Therefore, the effective sampling masks of $\mathbf{y}_{1}$ and $\mathbf{y}_{2}$ are:
    \begin{equation}
        \mathbf{M}_{y_{1} } = \mathbf{M}_{1} \mathbf{M}_{y},\,\
        \mathbf{M}_{y_{2} } = \mathbf{M}_{2} \mathbf{M}_{y},
    \label{eq:effective_mask}
    \end{equation}
which store the sampled locations in $\mathbf{y}_{1}$ and $\mathbf{y}_{2}$. The network performs reconstructions on $\mathbf{x}_{1}$ and $\mathbf{x}_{2}$, which are obtained by transforming $\mathbf{y}_{1}$ and $\mathbf{y}_{2}$ to the image domain using the multi-coil adjoint operator $\mathbf{A}^\mathrm{H}$, resulting in reconstructed images $\mathbf{x}_{1}^{\prime}$ and $\mathbf{x}_{2}^{\prime}$.

We propose a loss function consisting of two parts: image consistency loss $\mathcal{L}_{img}$ and cross k-space loss $\mathcal{L}_{ksp}$ as illustrated in Fig.~\ref{fig:Reconstruction_Network}(f):
\begin{equation}
    \mathcal{L}_{recon}=\mathcal{L}_{img}+\mathcal{L}_{ksp}.   
\label{eq:ReconLoss}
\end{equation}

The image consistency loss is defined as the mean squared error (MSE) between reconstructed images:
\begin{equation}
    \mathcal{L}_{img}= \left ( \mathbf{x}_{1}^{\prime}-\mathbf{x}_{2}^{\prime}  \right )^\mathrm{H}\left ( \mathbf{x}_{1}^{\prime}-\mathbf{x}_{2}^{\prime}  \right ).
\label{eq:ImgLoss}
\end{equation}
Since the re-undersampled input images $\mathbf{x}_{1}$ and $\mathbf{x}_{2}$ originate from the same k-space, the same reconstructed images $\mathbf{x}_{1}^{\prime}$ and $\mathbf{x}_{2}^{\prime}$ should be obtained. The image consistency loss thus forces the reconstructed images to be similar.

The cross k-space loss acts on the k-spaces $\mathbf{y}_{1}^{\prime}$ and $\mathbf{y}_{2}^{\prime}$ transformed from the reconstructed images $\mathbf{x}_{1}^{\prime}$ and $\mathbf{x}_{2}^{\prime}$. Data fidelity in the acquisition domain should be consistent and independent of the applied sampling mask. Thus each k-space is applied with the effective mask of the other image, i.e., $\mathbf{M}_{y_{2}}$ for $\mathbf{y}_{1}^{\prime}$, $\mathbf{M}_{y_{1}}$ for $\mathbf{y}_{2}^{\prime}$, yielding the k-spaces $\mathbf{y}_{1}^{\prime\prime}$ and $\mathbf{y}_{2}^{\prime\prime}$:
\begin{equation}
    \begin{matrix}
        y_{1}^{\prime\prime}=\mathbf{M}_{y_{2}}\mathbf{FS}(\mathbf{x}_{1}^{\prime}), \quad
        y_{2}^{\prime\prime}=\mathbf{M}_{y_{1}}\mathbf{FS}(\mathbf{x}_{2}^{\prime}).
    \end{matrix} 
    \label{eq:kspaceTransform}
\end{equation}
The cross k-space loss is defined as the mean absolute error (MAE) between the reconstructed k-space ($\mathbf{y}_{1}^{\prime\prime}, \mathbf{y}_{2}^{\prime\prime}$) and the true acquired samples of re-undersampled k-spaces ($\mathbf{y}_1,\mathbf{y}_2$):
\begin{equation}
\begin{split}
    \mathcal{L}_{ksp} &= \sqrt{(\mathbf{y}_{1}^{\prime\prime}-\mathbf{y}_{2})^\mathrm{H} (\mathbf{y}_{1}^{\prime\prime}-\mathbf{y}_{2})+\zeta }\\
    &+ \sqrt{(\mathbf{y}_{2}^{\prime\prime}-\mathbf{y}_{1})^\mathrm{H} (\mathbf{y}_{2}^{\prime\prime}-\mathbf{y}_{1})+\zeta },
\end{split}
\label{eq:KspLoss}
\end{equation}
where $\zeta$ is a small positive constant to prevent numerical instability. The cross k-space loss enforces the reconstructed k-space values to be close to the true values at the sampled points, thus driving the training to utilize the available sampling information to estimate missing k-space samples. We utilize MAE as k-space loss because of its robustness to outliers. It does not excessively penalize low-frequency signals, thus helping to preserve more image details.
\subsubsection{Network architecture}
The reconstruction network in Fig.~\ref{fig:Reconstruction_Network}(e) is a physics-based unrolled neural network with $n$ iterations, interleaved between UNet and DC layers. The architecture of UNet, the number of iterations, and DC layers are all the same as FE-Net to ensure consistency of the learned features. The complex-valued UNet shown in Fig.~\ref{fig:Reconstruction_Network}(c) has an additional residual connection between the input and the output. The encoder and decoder parts of each UNet contain two stages in which a 2D+t convolution is performed. Each 2D+t convolution consists of a 2D spatial convolution followed by a 1D temporal convolution, both utilizing ModReLU activation \cite{b36}. Separating the spatial and temporal convolutions reduces computational complexity and allows independent feature learning for each dimension \cite{b19}. Moreover, the in-between activation function increases nonlinearity, improving the fitting ability of the network.

To assist the network in reconstruction, the input images $\mathbf{x}_{1}$ and $\mathbf{x}_{2}$ will also undergo the pre-trained FE-Net. In each iteration, the extracted sampling-insensitive and subject-specific features will be concatenated with the features in the reconstruction network along the channel dimension before entering the bottleneck of each UNet, as illustrated in Fig.~\ref{fig:Reconstruction_Network}(e). This feature concatenation operation not only ensures a suitable initialization but also alleviates the interference of artifacts caused by highly undersampled datasets.

Data consistency layers are used to ensure data fidelity to the sampled k-space data. DC layers for the multi-coil dataset are implemented by gradient descent:
\begin{equation}
    \mathbf{x}_{dc,i}=\mathbf{x}_{i}-\lambda \mathbf{A}^\mathrm{H}(\mathbf{A}\mathbf{x}_{i}-\mathbf{y}), 
\label{eq:DClayer}
\end{equation}
where $\mathbf{x}_{i}$ and $\mathbf{x}_{dc,i}$ are the input and output images of the DC layer in the $i^{th}$ iteration, $\lambda$ is a trainable weighting parameter, and $\mathbf{y}$ is the acquired undersampled k-space. 

\section{Experiments}
\subsection{Dataset and sampling mask}
\subsubsection{Retrospective study}
The in-house 2D cardiac Cine dataset was acquired on a 1.5T MRI scanner (MAGNETOM Aera, Siemens Healthineers, Erlangen, Germany) using a balanced steady-state free precession (bSSFP) sequence. The sequence parameters are as follows: TE/TR=1.06/2.12 ms, flip angle=52°, bandwidth=915 Hz/px, spatial resolution=1.9 mm $\times $ 1.9 mm, slice thickness=8 mm, cardiac phases=25. The dataset contains 129 subjects, including 38 healthy subjects and 91 patients with heart diseases. We split it into 115 subjects for training, amongst which are 34 healthy volunteers and 81 patients. The remainder of the dataset was used for testing. Both FE-Net and the reconstruction network were trained on the same dataset. The local ethics committee approved the study, and all subjects gave written consent. 

This study utilizes a variable density incoherent spatiotemporal acquisition (VISTA) \cite{b37} undersampling for both feature learning (first step) and reconstruction (second step). For FE-Net training, acceleration factors are randomly selected from $2\times$ to $16\times$ for both contrastive learning and the information maximization method. Notably, the initial images to create positive pairs for training are $2\times$ conventional parallel imaging undersampled data. In the second reconstruction step, the initial undersampled k-space is re-undersampled for training with masks $\mathbf{M}_{1}$ and $\mathbf{M}_{2}$. Initial undersampling is obtained with a fixed $2\times$ VISTA sampling to reflect a realistic simulation of the prospective case. Two re-undersampling masks are independent and have random acceleration rates between $2\times$ to $16\times$ for each training step, resulting in an effective acceleration rate of $2\times$ to $28\times$ per mask. Following previous work \cite{b48}, the re-undersampling masks should conform to similar sampling characteristics. In our study, we thus keep the VISTA sampling distribution for initial undersampling and re-undersampling to train the reconstruction network. It is important to note that the initial undersampling and re-undersampling strategy is only used during the training phase. In inference, singly undersampled images are used as inputs (Fig.~\ref{fig:Reconstruction_Network}(d)). In the following, the acceleration $R$ refers to the effective acceleration rate.

\subsubsection{Prospective study}
We further tested the pre-trained models on the prospectively undersampled data from the OCMR dataset \cite{ocmr}. The prospective data were collected in real-time on a 1.5T MRI scanner (MAGNETOM Avanto, Siemens Healthineers, Erlangen, Germany) using an SSFP sequence under free-breathing with temporal resolution=43.4 ms, spatial resolution=2.21 mm $\times $ 2.46 mm, and slice thickness=8 mm. To validate the generalization performance of each method, we did not conduct any training or fine-tuning on the prospective data. The pre-trained models were directly used to reconstruct the prospectively undersampled data.

\subsection{Implementation details}
Both FE-Net and the reconstruction network operate in the complex domain, except for the MLPs in the feature learning network, for which the data are converted to two-channel real numbers. The convolutional kernel size is 5 for spatial and 3 for temporal filters. The number of kernels is annotated beside each layer in Fig.~\ref{fig:Reconstruction_Network}(c). Hyperparameters are set to $\lambda=\mu=25$ and $v=1$ in (\ref{eq:VICReg}), $\epsilon=0.0001$ in (\ref{eq:std}), $\zeta=1\times$10$^{-9}$ in (\ref{eq:KspLoss}). We initilized $\lambda$ in (\ref{eq:DClayer}) as 1. Both feature learning and reconstruction networks iterate $n=3$ times.

We implemented the proposed SSFL-Recon framework using \textit{Tensorflow} v2.6.0 with \textit{Keras} v2.6.0. Complex-valued operations such as convolutions and activations are implemented by MERLIN v0.3 \cite{b38}. Networks were trained using an Adam optimizer \cite{b39} with a learning rate of 4$\times10^{-4}$, 30 epochs were trained for self-supervised feature learning, and 200 epochs were trained for the reconstruction step. The source code is publicly available: \href{https://github.com/midas-tum/SSFL-Recon}{https://github.com/midas-tum/SSFL-Recon}.

\subsection{Experimental settings}
We compare the proposed SSFL-Recon framework to a conventional sparse and low-rank based method: k-t SLR \cite{b40}, two SSL reconstruction networks, SSDU \cite{b22} and PARCEL \cite{b33}, and a supervised training of the same architecture as our reconstruction network using fully-sampled images as ground-truths. For traditional methods, parameters were adjusted to achieve optimal performance. For other experiments, we used the same network architecture and sampling strategy (including initial and re-undersampling) to make a fair comparison between learning strategies. The total trainable parameters, training times, and inference times of different methods for a Cine image sequence are summarized in Tab.~\ref{tab:summary_comparison}.

\begin{table}[!t]
\centering
\caption{Summary of trainable parameters, training times (hour), and reconstruction times (second) for a complete 2D cardiac Cine data in all examined algorithms.}
\label{tab:summary_comparison}
\resizebox{\columnwidth}{!}{%
\begin{tabular}{cc|c|c|c}
\toprule
\multicolumn{2}{c|}{Methods}                                              & Trainable Params  & Training(h) & Recon(s) \\ \midrule
\multicolumn{1}{c|}{Conventional}                         & kt-SLR        & None              & None              & 646.80         \\ \midrule
\multicolumn{1}{c|}{\multirow{6}{*}{\begin{tabular}[c]{@{}c@{}}Deep learning-\\ based methods\end{tabular}}} & PARCEL        & 30,950,111        & 55                & 6.563          \\
\multicolumn{1}{c|}{}                                     & SSDU          & 417,358           & 36                & 5.266          \\
\multicolumn{1}{c|}{}                                     & Supervised    & 417,358           & 25                & 4.820          \\
\multicolumn{1}{c|}{}                                     & SSL-Recon     & 417,358           & 68                & 5.890          \\
\multicolumn{1}{c|}{}                                     & SSFL-Recon(c) & 320,432 + 417,358 & 7 + 96            & 6.442          \\
\multicolumn{1}{c|}{}                                     & SSFL-Recon(v) & 320,432 + 417,358 & 9 + 96            & 6.443          \\ \bottomrule
\end{tabular}
}
\end{table}

To evaluate the role of feature learning, we conducted an ablation study. In the ablation experiments, we adopted the same self-supervised learning strategy to train the reconstruction network but without FE-Net and the assistance of pre-learned features, in the following named as SSL-Recon.

In the results, we tested only up to $16\times$ when comparing different methods because we observed stable performance of most networks within this range. However, when showcasing the performance of the proposed SSFL-Recon framework, we pushed the acceleration to $21\times$, which did not exhibit significant reconstruction errors. In the quantitative evaluations, we calculated the normalized root mean squared error (NRMSE), peak signal-to-noise ratio (PSNR), and structural similarity index measure (SSIM) for all slices across all subjects in the test dataset.

\section{Results}
\subsection{Reconstructions of SSFL-Recon framework}
\begin{figure*}[!t]
\centerline{\includegraphics[width=\textwidth]{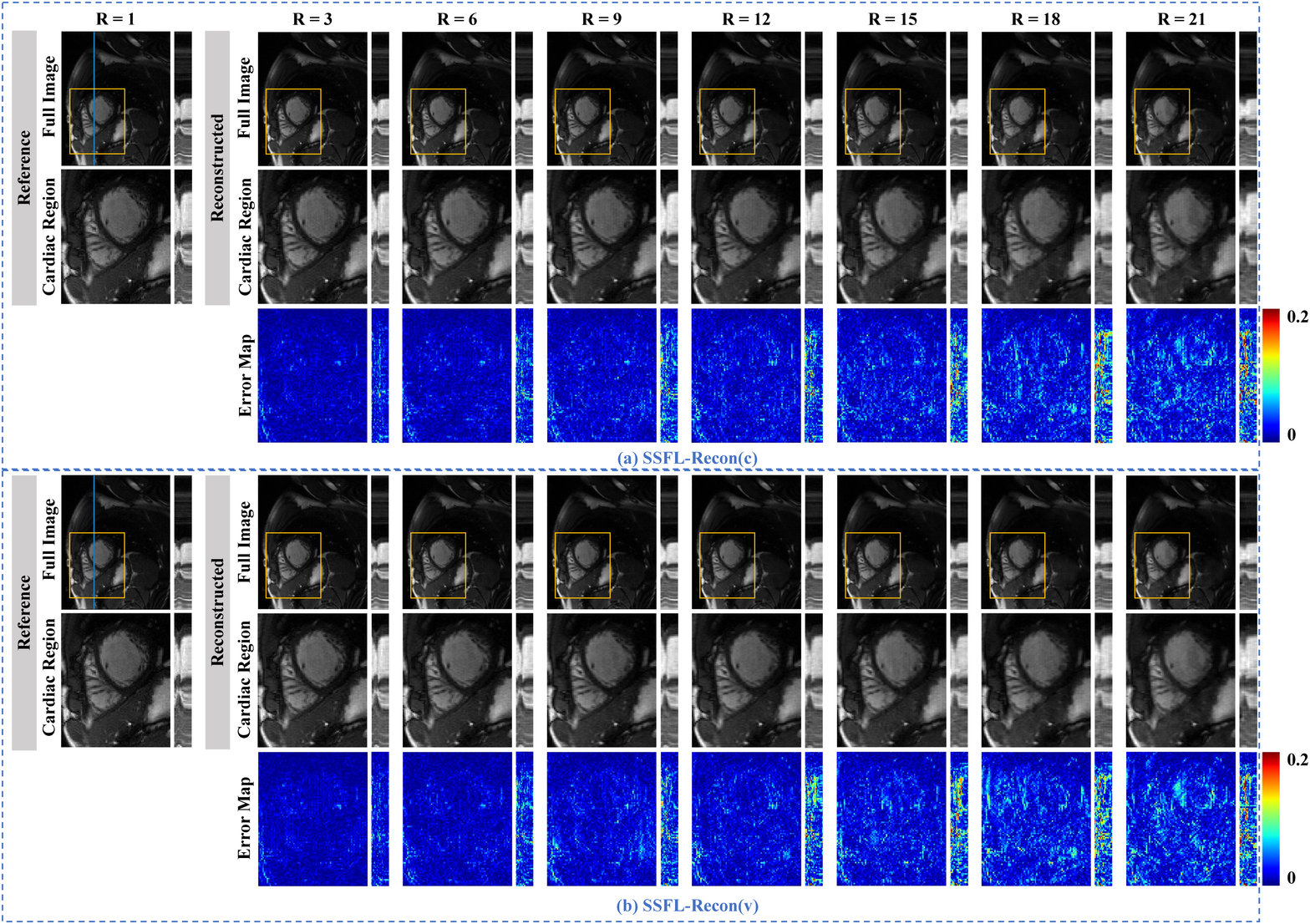}}
\caption{Reconstructions in spatial (x-y) and spatio-temporal (y-t) plane of the proposed (a) SSFL-Recon(c) and (b) SSFL-Recon(v) using a retrospectively VISTA undersampling in a patient with transposition of the great vessels. Results for the acceleration factors $R=3, 6, 9, 12, 15, 18, 21$ are shown in each column. The dynamic performance in the y-t plane corresponds to the blue line in the reference x-y plane image. The second row of each sub-figure shows the enlarged views of the cardiac region (yellow box region). The third row presents the corresponding 5-times scaled relative error maps between the reconstructed images and the fully-sampled reference.} 
\label{fig:Recon_results_diff_R}
\end{figure*}

Fig.~\ref{fig:Recon_results_diff_R} presents the reconstructed spatial (x-y) and spatio-temporal (y-t) 2D cardiac Cine images, along with magnified views of the cardiac region and corresponding relative error maps for the proposed SSFL-Recon(c) and SSFL-Recon(v). We chose a patient with transposition of the great vessels from the test dataset and depicted the reconstruction results for different accelerations. As the acceleration factors were random during training, the trained network can adapt to various accelerations during inference.

Results demonstrate that the proposed method can perform high-quality reconstructions under different accelerations. The reconstructed images are artifact-free up to $18\times$ acceleration. Even under $21\times$ acceleration factor, the 5-times enhanced error map shows only a small residual aliasing in the heart region. Nevertheless, myocardial delineation and small structures like the papillary muscles are still well depicted. We observe consistent performance throughout the whole test dataset. The SSFL-Recon framework with either contrastive feature learning or information maximization method using VICReg loss presents convincing performance for different acceleration rates. As shown in Fig.~\ref{fig:quant_plot}, the proposed SSFL-Recon framework achieves a high average SSIM score of nearly 0.88 calculated on the whole test dataset under $16\times$ acceleration. The difference between the two feature learning methods is not apparent. We observe similar performance for both networks, which is also indicated in Table~\ref{tab:ablation_quant}.

\subsection{Comparison studies}
\subsubsection{Retrospective study}
\begin{figure*}[!t]
\centerline{\includegraphics[width=\textwidth]{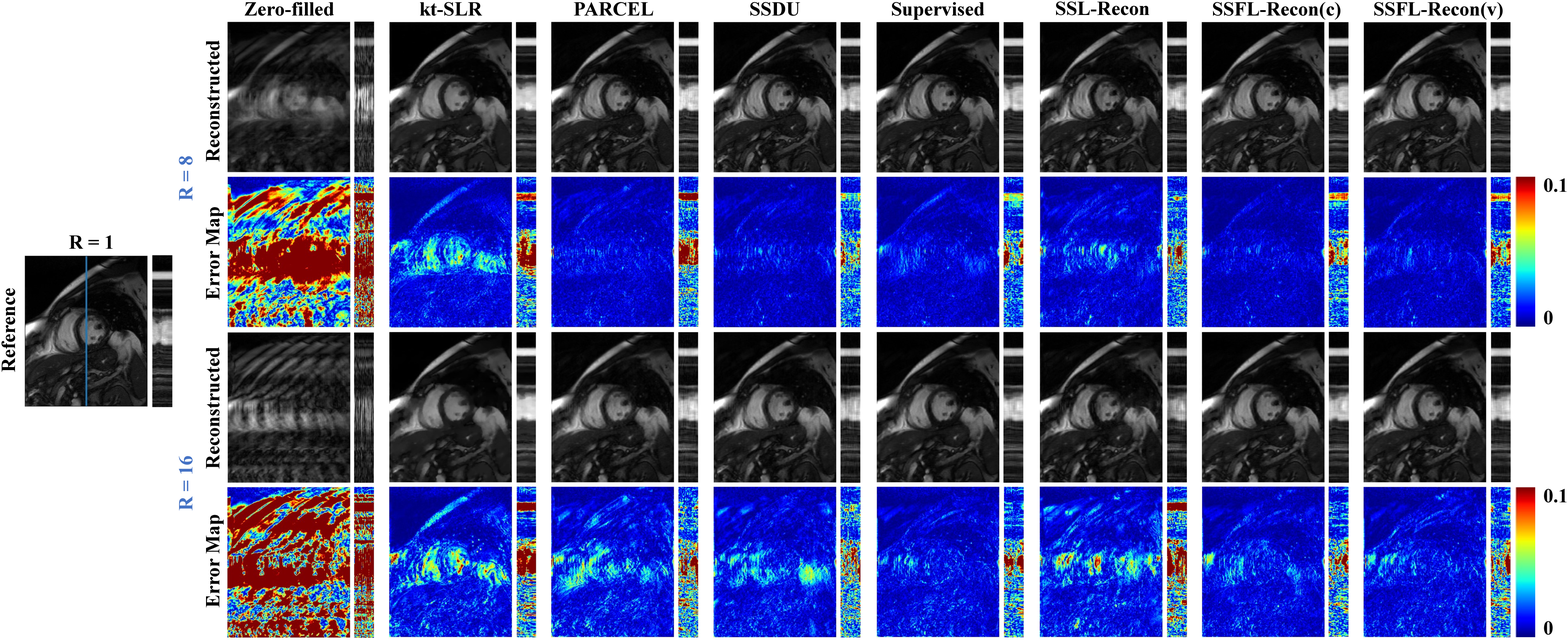}}
\caption{Reconstructions in spatial (x-y) and spatio-temporal (y-t) plane of the proposed SSFL-Recon frameworks in comparison to zero-filled, kt-SLR \cite{b40}, PARCEL \cite{b33}, SSDU \cite{b22}, supervised learning, and the ablation study SSL-Recon for a healthy subject who was retrospectively undersampled with VISTA sampling. The dynamic performance in the y-t plane corresponds to the blue line in the reference x-y plane image. Both $R=8$ (top) and $R=16$ (bottom) reconstructions are shown alongside the corresponding relative error maps.} 
\label{fig:compare_healthy}
\end{figure*}

Fig.~\ref{fig:compare_healthy} presents the qualitative comparison between kt-SLR \cite{b40}, PARCEL \cite{b33}, SSDU \cite{b22}, supervised learning of the investigated architecture, and the proposed SSFL-Recon framework for a healthy test subject with retrospectively $8\times$ and $16\times$ undersamplings. The reconstructions of the traditional kt-SLR method are acceptable but exhibit an over-smoothing effect, leading to blurred edges and lost details such as papillary muscles. PARCEL performs well in reconstructing the $8\times$ accelerated image. However, the error map becomes more pronounced as the acceleration factor increases to $16\times$. Residual artifacts can be observed in the interventricular septum region. Another self-supervised reconstruction method, SSDU, performs similarly to PARCEL. Notable artifacts are observed in the reconstructed left and right ventricle under $16\times$ acceleration. In contrast, the proposed SSFL-Recon frameworks achieve high-quality reconstructions for both moderately $8\times$ accelerated and highly $16\times$ accelerated images. Both SSFL-Recon methods with different feature learning approaches outperform PARCEL and SSDU, demonstrating comparable performance to supervised learning, even exhibiting better intensity reconstruction at $8\times$ acceleration.

\begin{figure*}[!t]
\centerline{\includegraphics[width=\textwidth]{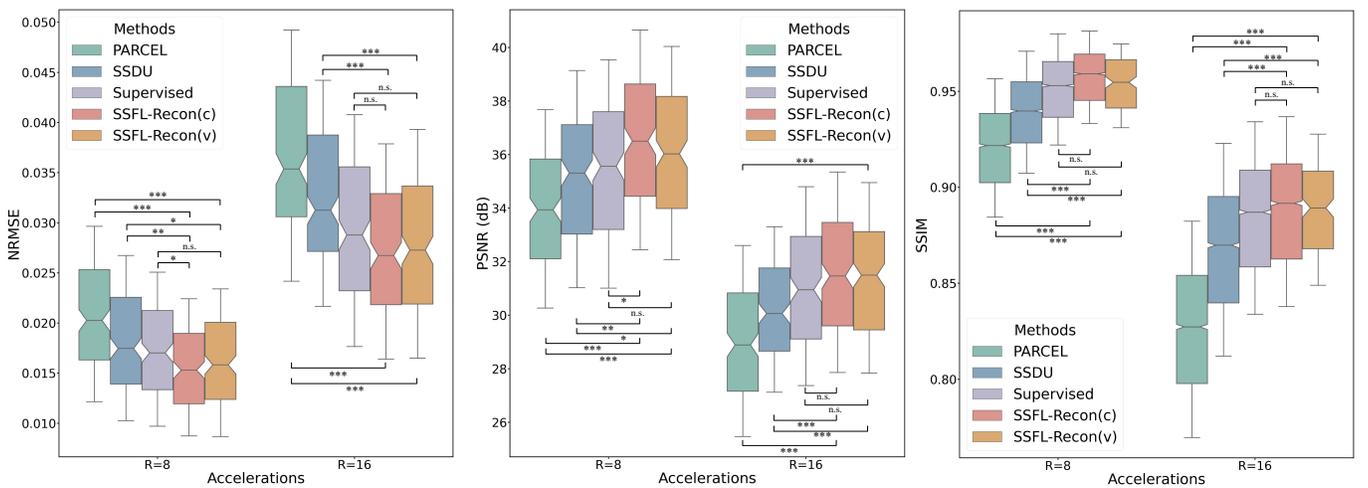}}
\caption{Quantitative comparison in terms of NRMSE, PSNR, and SSIM between PARCEL \cite{b33}, SSDU \cite{b22}, supervised learning, and the proposed SSFL-Recon frameworks. Each metric was computed for accelerations $R=8$ and $R=16$. Results are calculated for all subjects in the test dataset and depicted in box plots (horizontal line: median, box: 25\% and 75\% percentile, whiskers: 0.5 $\times$ interquartile range). Asterisk denotes statistically significant differences calculated by Wilcoxon rank sum test: \text{*}: $p<0.05$; \text{**}: $p<0.01$; \text{***}: $p<0.001$; n.s: not significant.}
\label{fig:quant_plot}
\end{figure*}

The dynamic information in the spatio-temporal (y-t) images illustrates that kt-SLR exhibits the same over-smoothing characteristics as observed in the spatial (x-y) plane, leading to obvious error maps. Vertical line artifacts remain in PARCEL and SSDU under $16\times$ acceleration, i.e., inconsistent brightness and contrast throughout the cardiac cycle were recovered. Conversely, our proposed SSFL-Recon can effectively reconstruct dynamic behavior with high fidelity compared to the fully-sampled reference, which holds significant value for the assessment of cardiac function.

For a fair and comprehensive evaluation of DL-based self-supervised reconstruction methods, we computed NRMSE, PSNR, and SSIM over all slices of all test subjects under $8\times$ and $16\times$ retrospectively VISTA undersamplings. The results are summarized as box plots shown in Fig.~\ref{fig:quant_plot}. Although in Fig.~\ref{fig:compare_healthy}, different DL-based reconstruction methods exhibit similar performance for the selected slice under $8\times$ acceleration, a more pronounced difference is observed when qualitatively evaluating all slices in the test dataset, as shown in Fig.~\ref{fig:quant_plot}. At $8\times$ acceleration, both proposed SSFL-Recon frameworks outperform PARCEL and SSDU in all metrics. In terms of SSIM, which is visually closer to human eye perception, the average values of SSFL-Recon(c) and SSFL-Recon(v) are all above 0.95, which is better than the average of PARCEL (91.7\%) and SSDU (93.5\%). This advantage further expands with increased acceleration. At $16\times$ acceleration, the proposed SSFL-Recon methods not only outperform PARCEL and SSDU with noticeable differences but also demonstrate comparable performance to supervised learning, even exhibiting superior performance in terms of PSNR and SSIM as indicated in Table~\ref{tab:ablation_quant}.

\subsubsection{Prospective study}
\begin{figure}[!t]
\centerline{\includegraphics[width=\columnwidth]{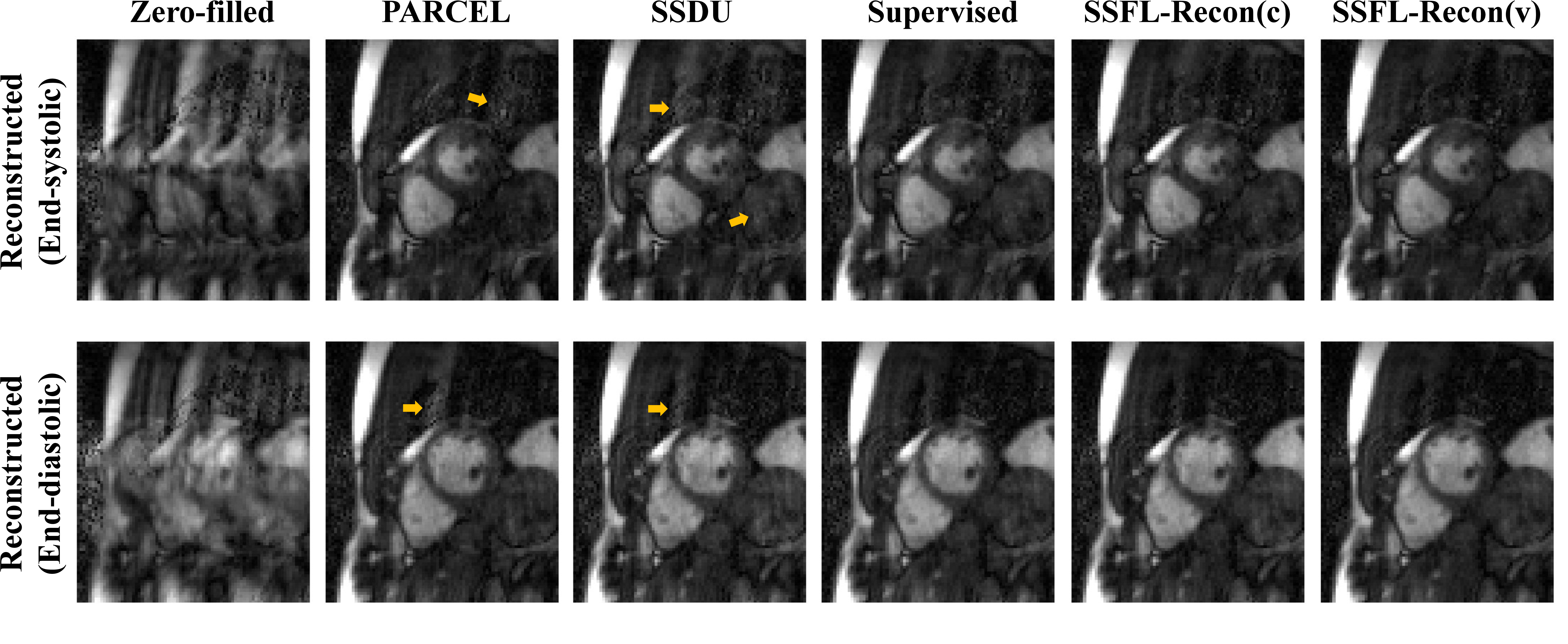}}
\caption{Reconstructions in spatial (x-y) plane of the proposed SSFL-Recon frameworks in comparison to zero-filled, PARCEL \cite{b33}, SSDU \cite{b22}, and supervised learning for a healthy subject of the OCMR dataset, which was prospectively undersampled with VISTA sampling (R=10). End-systolic and end-diastolic phases are shown in the top and bottom rows. Yellow arrows point out some obvious residual aliasing.}
\label{fig:prosp_compare}
\end{figure}

Fig.~\ref{fig:prosp_compare} displays the reconstruction results of a prospectively $10\times$ undersampled short-axis slice from the OCMR dataset \cite{ocmr}. We compared the performance of deep learning-based methods: PARCEL \cite{b33}, SSDU \cite{b22}, supervised learning, and the proposed SSFL-Recon frameworks for end-systolic and end-diastolic phases. As mentioned before, the prospective study aims to investigate the generalization ability of the method. Therefore, the pre-trained networks were applied directly for reconstruction without any fine-tuning. As a result, for most methods, the reconstructed image quality decayed compared to the retrospective study. In both end-systolic and end-diastolic phases, noticeable aliasing artifacts remain in the reconstructed images from PARCEL and SSDU (as indicated by the yellow arrows). In contrast, the proposed SSFL-Recon frameworks demonstrate reconstruction quality comparable to the retrospective setting and to the supervised learning, with more apparent structural details and improved contrast in the myocardial region.

\subsection{Ablation studies}

\begin{figure}[!t]
\centerline{\includegraphics[width=\columnwidth]{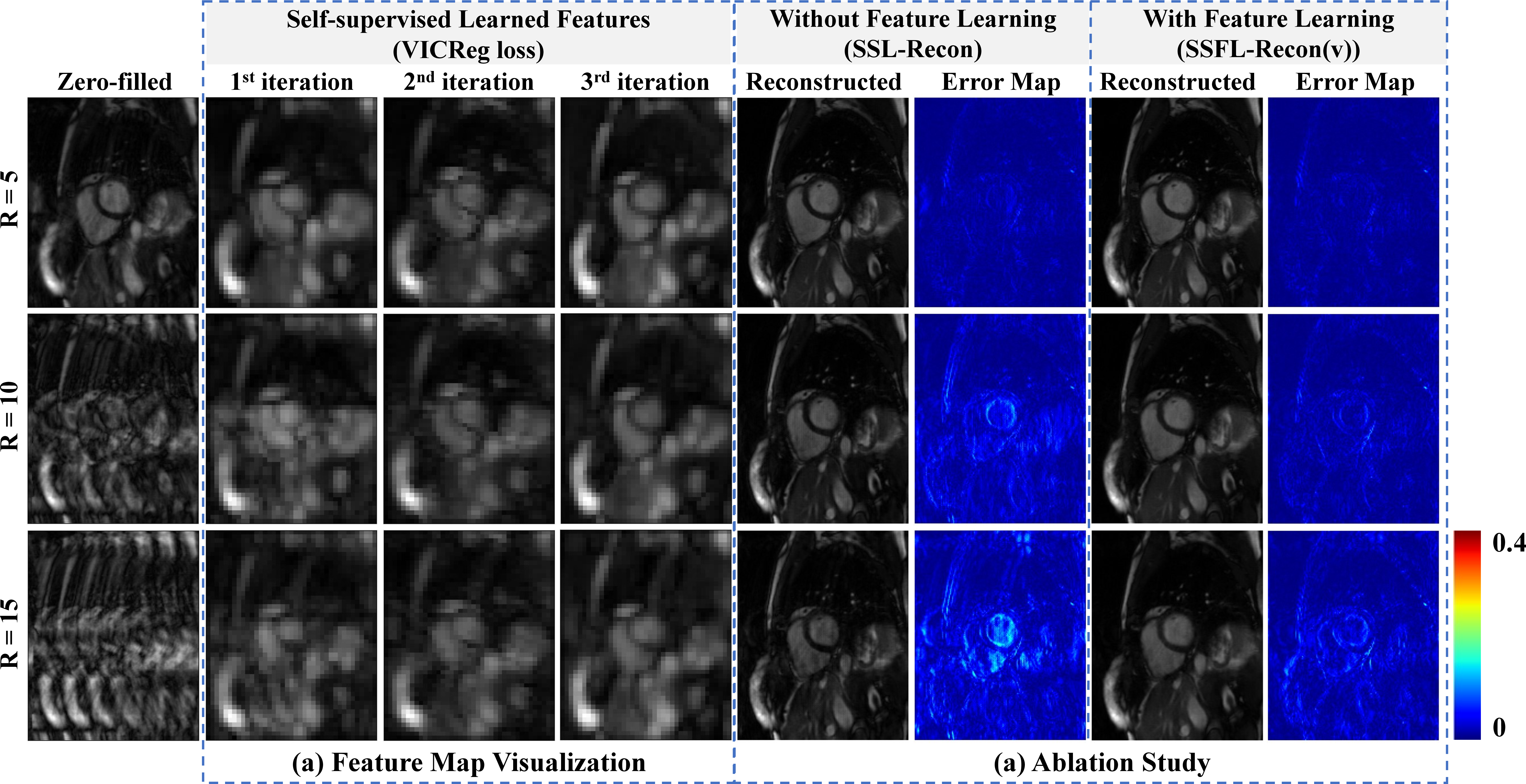}}
\caption{(a) The learned features in the first step of SSFL-Recon(v). Each column presents the features at different iterations in the unrolled network under $R=5,10,15$. (b) Ablation study: the reconstruction results of SSL-Recon compared to SSFL-Recon(v) under $R=5,10,15$.}
\label{fig:ablation_vicreg}
\end{figure}

\begin{figure}[!t]
\centerline{\includegraphics[width=\columnwidth]{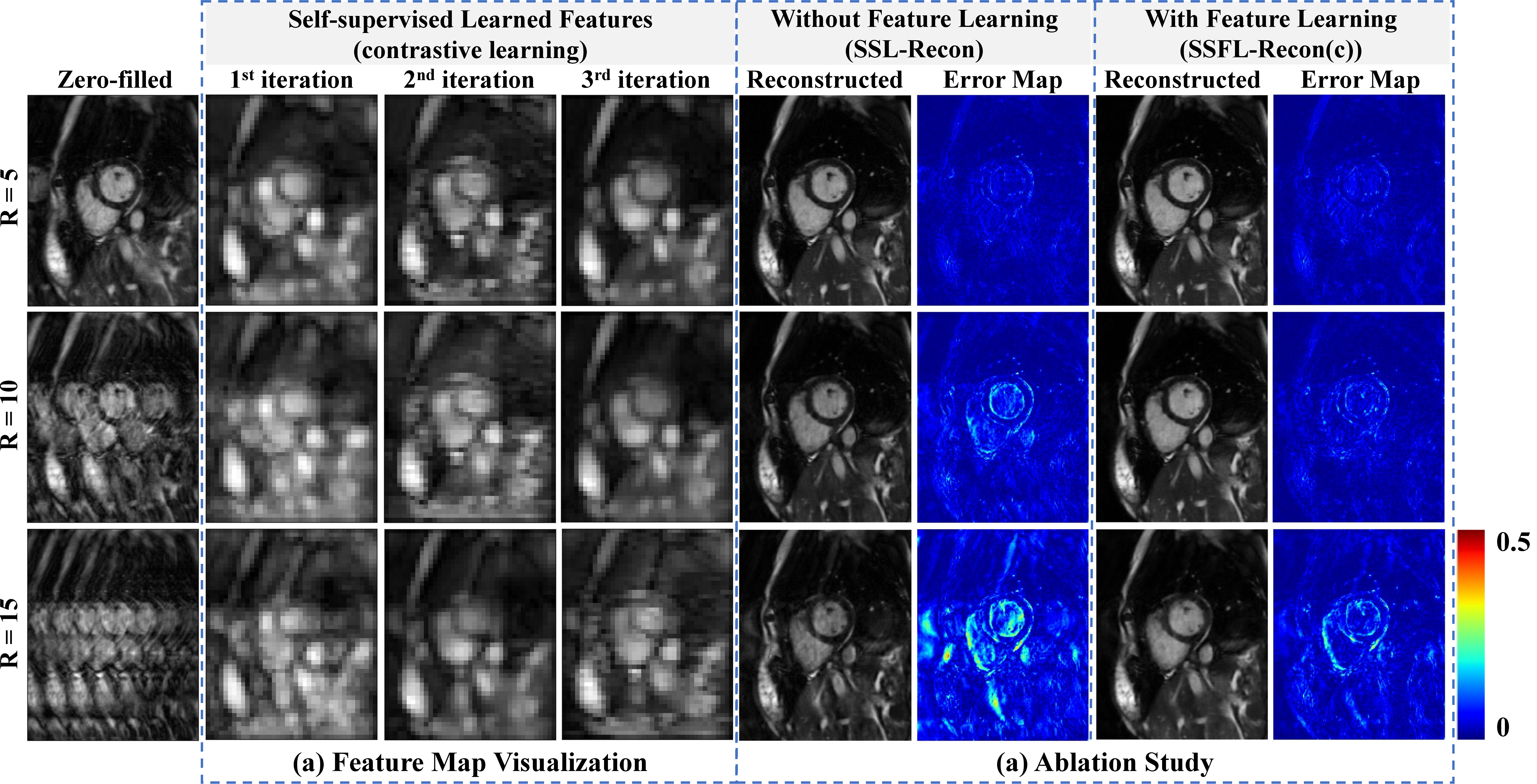}}
\caption{(a) The learned features in the first step of SSFL-Recon(c). Each column presents the features at different iterations in the unrolled network under $R=5,10,15$. (b) Ablation study: the reconstruction results of SSL-Recon compared to SSFL-Recon(c) under $R=5,10,15$.}
\label{fig:ablation_contrastive}
\end{figure}

In Fig.~\ref{fig:ablation_vicreg}(a), we present the learned features in each iteration of FE-Net using VICReg loss for accelerations $5\times$, $10\times$, and $15\times$. We observe that FE-Net in the first step of the proposed SSFL-Recon(v) framework can learn the cardiac anatomy as a low-pass filtered clean image from the undersampled dataset. The feature extraction ability is consistent and reliable across different noise and undersampling levels. A similar performance can be observed in Fig.~\ref{fig:ablation_contrastive}(a), which shows the extracted features of a different slice from another subject learned by contrastive learning.

It is worth noting that, as the learned features are the deep-level feature maps encoded by the encoder, the acquired features contain more global information, specifically the overall cardiac structure. The clear anatomical guidance aids the network in effectively removing artifacts, particularly in scenarios where aliasing artifacts are severe at high acceleration rates.

In Fig.~\ref{fig:compare_healthy}, we additionally present the reconstructed images by SSL-Recon, which is the ablated experiment without the feature learning step. In Fig.~\ref{fig:ablation_vicreg}(b), we show the reconstructions of SSFL-Recon(v) and SSL-Recon under $5\times$, $10\times$, and $15\times$ accelerations. Compared to the proposed method with the feature learning step, the reconstruction results of the self-supervised reconstruction network without FE-Net show significantly more residual aliasing artifacts under high accelerations. We observe a similar performance by comparing SSFL-Recon(c) and SSL-Recon shown in Fig.~\ref{fig:ablation_contrastive}(b) for another slice. The superior performance demonstrates the effectiveness of the proposed feature learning strategy.

To evaluate the performance over the whole test dataset, we computed NRMSE, PSNR, and SSIM under $8\times$ and $16\times$ accelerations. Results are summarized in Table~\ref{tab:ablation_quant}. The quantitative analysis also reflects the advantages of the feature learning step. At the lower acceleration factor of $8\times$, both SSFL-Recon methods exhibit high mean SSIM values surpassing 0.95, indicating a high resemblance to the fully-sampled reference images. In contrast, SSL-Recon without feature learning yields an average SSIM value below 0.9. As the acceleration increases to $16\times$, both SSFL-Recon methods demonstrate high mean PSNR values exceeding 30dB and structural similarity close to 0.9, outperforming SSL-Recon. Furthermore, the proposed SSFL-Recon framework displays smaller variances, implying enhanced robustness and consistent reconstruction performance across various test samples.

\begin{table}
\caption{Quantitative evaluation of SSL-Recon, SSFL-Recon(c), SSFL-Recon(v), and the supervised learning: The average and standard deviation of NRMSE, PSNR, and SSIM in the reconstructed images for all subjects in the test dataset at $R=8$ and $R=16$ (mean±std). The asterisk(*) denotes statistically significant differences (p $<$ 0.05.) between each method with the best performance (indicated in bold), as calculated by the Wilcoxon rank sum test.}
\label{tab:ablation_quant}
\resizebox{\columnwidth}{!}{%
\begin{tabular}{c|cccc}
\toprule
Accelerations               & Methods         & NRMSE                 & PSNR (dB)              & SSIM                    \\
\midrule
\multirow{3}{*}{8$\times$}  & SSL-Recon        & 0.029±0.009$^{*}$     & 30.992±2.670$^{*}$     & 0.892±0.039$^{*}$       \\
                            & SSFL-Recon(c)   & \textbf{0.016±0.005}  & \textbf{36.325±2.836}  & \textbf{0.954±0.022}    \\
                            & SSFL-Recon(v)   & 0.017±0.005           & 35.892±2.838           & 0.951±0.019             \\
                            & Supervised      & 0.018±0.006$^{*}$     & 35.326±2.855$^{*}$     & 0.947±0.025       \\
\midrule
\multirow{3}{*}{16$\times$} & SSL-Recon        & 0.056±0.013$^{*}$     & 25.371±2.333$^{*}$     & 0.769±0.053$^{*}$       \\
                            & SSFL-Recon(c)   & 0.029±0.010          & \textbf{31.344±2.785}           & 0.884±0.041       \\
                            & SSFL-Recon(v)   & \textbf{0.029±0.009}  & 31.229±2.660  & \textbf{0.886±0.031}   \\
                            & Supervised      & 0.030±0.010           & 30.955±2.677           & 0.881±0.040   \\
\bottomrule
\end{tabular}%
}
\end{table}

\section{Discussion}
In this work, we developed a novel self-supervised feature learning assisted reconstruction framework (SSFL-Recon) and demonstrated its potential to reconstruct cardiac Cine images if only undersampled images are available during training. The SSFL-Recon framework consists of two steps: (i) self-supervised feature learning and (ii) self-supervised reconstruction. We proposed a feature learning strategy that enables the network to learn sampling-insensitive and subject-specific features from undersampled images. With the assistance of the learned cardiac structure features in the first step, the reconstruction network exhibits improved performance in artifact removal at high acceleration rates. In the second step of the SSFL-Recon framework, a customized self-supervised reconstruction loss is designed, encompassing image consistency and cross k-space loss. The utilization of information in both image and k-space domains ensures the maximized sharing of multi-domain information.

\subsection{Feature learning strategy}
Both feature learning methods demonstrate the feasibility of learning subject-specific and sampling-insensitive features directly from undersampled data, thereby enhancing the generalization performance of the reconstruction network across different undersampling masks (of similar sampling distribution). In the ablation study, we visualized the learned features and demonstrated the advantages brought by FE-Net in the reconstruction process. Most existing self-supervised reconstruction methods and the ablated SSL-Recon demonstrate stable performance within the range of $16\times$ acceleration. However, with FE-Net, the acceleration can be further pushed to $21\times$, as shown in Fig.~\ref{fig:Recon_results_diff_R}, potentially utilizing highly undersampled data for training. 

When comparing the two feature learning strategies, we observed similar reconstruction performance. However, compared to contrastive learning, we recommend using the information maximization method (such as VICReg), considering that contrastive learning requires negative pairs in addition to positive pairs during training. The demand for negative pairs adds extra memory burden and computational complexity. Moreover, the performance of contrastive learning depends on the number of negative pairs, often requiring a larger number for better feature learning, which poses a significant challenge for high-dimensional dynamic MRI data. The information maximization method avoids the need for negative pairs, achieving comparable results with less GPU consumption and computational burden.

Additionally, we observed that the proposed feature learning strategy demonstrates strong transferability. Specifically, we conducted experiments by splitting the original dataset into disjoint subsets, using one subset for feature learning and the other for the reconstruction step. Both quantitative and qualitative results indicated comparable performance to the scenario where the same dataset was used for both steps. These findings suggest that the feature extractor can be effectively trained on subsets with similar data distributions, enhancing its potential applicability when only limited, similar datasets are available.

The proposed feature learning strategy is theoretically applicable to different sampling trajectories and masks. In this work, we specifically focused on VISTA sampling \cite{b37} due to the investigation of cardiac Cine MRI. Considering the nature of this undersampling technique, which is applied in the y-t plane, the network still incorporates temporal convolutions despite the feature learning loss in the first step being applied in the 2D spatial domain.

\subsection{Loss function of self-supervised reconstruction}
In the second step of the proposed SSFL-Recon framework, we employed a self-defined loss function (\ref{eq:ReconLoss}) comprising two components: image consistency loss and cross k-space loss. To investigate the contributions of these two losses, we conducted experiments using a single loss, either image consistency loss or cross k-space loss. We observed that the network using only image consistency loss failed completely to reconstruct images. This finding aligns with observations in DDSS \cite{b24}, where the network fails to converge when using only an appearance consistency loss in the image domain. In contrast, when using only the cross k-space loss, the network can converge, and further improvement was achieved upon incorporating the image consistency loss.

\subsection{Limitations and outlook}
We acknowledge the following limitations. The global representations obtained by FE-Net are insufficient to represent fine details. We will investigate local feature learning strategies with more data augmentation and the integration of learned features for different architectures in the future. Currently, we focus on Cartesian sampling, which is commonly used in clinical cardiac Cine MRI. The proposed method is theoretically applicable to other sampling strategies, such as non-Cartesian trajectories, but further validation is needed in future work. In addition, the re-undersampling is entirely random, aiming to introduce more randomness during training and to enhance the generalization capabilities. A learning of the optimal initial or re-undersampling pattern \cite{bahadir2019learning, zibetti2021fast} might yield further improvements and warrants investigation in future work. While existing experiments have demonstrated the feasibility of the proposed method for cardiac cine MRI, the concept is generally applicable to other imaging applications, body regions, and dimensionalities, such as 2D datasets from the fastMRI. We plan to explore the generalizability of these domain shifts for the proposed method in future work. Furthermore, we plan to deploy the framework on the scanner to explore the effectiveness of the proposed methods on prospectively undersampled data with higher acceleration rates.

\section{Conclusion}
In this work, we propose a novel self-supervised feature learning assisted reconstruction framework (SSFL-Recon) for dynamic cardiac Cine data that would allow the reconstruction using only undersampled data. We pre-train a feature extractor to learn subject-specific and sampling-insensitive features from undersampled data as the first step. The extracted features are incorporated into a self-supervised reconstruction network and assist in removing aliasing artifacts. The physics-based unrolled network can learn reconstructions from only undersampled k-space through self-defined image consistency and cross k-space losses. Both qualitative and quantitative results demonstrate that the proposed SSFL-Recon framework outperforms other self-supervised reconstruction methods, achieving comparable, even better performance to supervised learning.


\begin{thebibliography}{00}

\bibitem{b1} K. P. Pruessmann, M. Weiger, M. B. Scheidegger, and P. Boesiger, ``SENSE: sensitivity encoding for fast MRI,'' \emph{Magn. Reson. Med.}, vol. 42, no. 5, pp. 952--962, Nov. 1999.

\bibitem{b2} M. A. Griswold \emph{et al., ``}Generalized autocalibrating partially parallel acquisitions (GRAPPA),'' \emph{Magn. Reson. Med.}, vol. 47, no. 6, pp. 1202--1210, Jun. 2002.

\bibitem{b3} M. Lustig, and J. M. Pauly, ``SPIRiT: iterative self‐consistent parallel imaging reconstruction from arbitrary k‐space,'' \emph{Magn. Reson. Med.}, vol. 64, no. 2, pp. 457--471, Aug. 2010.

\bibitem{b4} M. Uecker \emph{et al., ``}ESPIRiT—an eigenvalue approach to autocalibrating parallel MRI: where SENSE meets GRAPPA.,'' \emph{Magn. Reson. Med.}, vol. 71, no. 3, pp. 990--1001, Mar. 2014.

\bibitem{b5} L. Cha{\^a}ri, J. C. Pesquet, A. Benazza-Benyahia, and P. Ciuciu, ``A wavelet-based regularized reconstruction algorithm for SENSE parallel MRI with applications to neuroimaging,'' \emph{Med. Image Anal.}, vol. 15, no. 2, pp. 185--201, Apr. 2011.

\bibitem{b6} S. Osher, M. Burger, D. Goldfarb, J. Xu, and W. Yin, ``An iterative regularization method for total variation-based image restoration,'' \emph{Multiscale Model. Simul.}, vol. 4, no. 2, pp. 460--489, 2005.

\bibitem{b7} K. T. Block, M. Uecker, and J. Frahm, ``Undersampled radial MRI with multiple coils. Iterative image reconstruction using a total variation constraint,'' \emph{Magn. Reson. Med.}, vol. 57, no. 6, pp. 1086--1098, Jun. 2007.

\bibitem{b8} Q. Liu, S. Wang, L. Ying, X. Peng, Y. Zhu, and D. Liang, ``Adaptive dictionary learning in sparse gradient domain for image recovery,'' \emph{IEEE Trans. Image Process.}, vol. 22, no. 12, pp. 4652--4663, Aug. 2013.

\bibitem{b9} J. Caballero, A. N. Price, D. Rueckert, and J. V. Hajnal, ``Dictionary learning and time sparsity for dynamic MR data reconstruction,'' \emph{IEEE Trans. Med. Imaging}, vol. 33, no. 4, pp. 979--994, Jan. 2014.

\bibitem{b10} D. Weller, ``Reconstruction with dictionary learning for accelerated parallel magnetic resonance imaging,'' \emph{IEEE SSIAI}, Santa Fe, NM, USA, pp. 105--108, Mar. 2016.

\bibitem{b11} B. Liu, Y. M. Zou, and L. Ying, ``SparseSENSE: application of compressed sensing in parallel MRI,'' \emph{IEEE EMBS Int. Conf. ITAB}, Shenzhen, China, pp. 127--130, May. 2008.

\bibitem{b12} D. Lee, K. H. Jin, E. Y. Kim, S. H. Park, and J. C. Ye, ``Acceleration of MR parameter mapping using annihilating filter‐based low rank hankel matrix (ALOHA),'' \emph{Magn. Reson. Med.}, vol. 76, no. 6, pp. 1848--1864, Dec. 2016.

\bibitem{b13} M. Ak{\c{c}}akaya, S. Moeller, S. Weing{\"a}rtner, and K. U{\u{g}}urbil, ``Scan‐specific robust artificial‐neural‐networks for k‐space interpolation (RAKI) reconstruction: Database‐free deep learning for fast imaging,'' \emph{Magn. Reson. Med.}, vol. 81, no. 1, pp. 439--453, Jan. 2019.

\bibitem{b14} D. Lee, J. Yoo, S. Tak, and J. C. Ye, ``Deep residual learning for accelerated MRI using magnitude and phase networks,'' \emph{IEEE Trans. Biomed. Eng.}, vol. 65, no. 9, pp. 1985--1995, Apr. 2018.

\bibitem{b15} A. Hauptmann, S. Arridge, F. Lucka, V. Muthurangu, and J. A. Steeden, ``Real‐time cardiovascular MR with spatio‐temporal artifact suppression using deep learning–proof of concept in congenital heart disease,'' \emph{Magn. Reson. Med.}, vol. 81, no. 2, pp. 1143--1156, Feb. 2019.

\bibitem{b16} A. Kofler, M. Dewey, T. Schaeffter, C. Wald, and C. Kolbitsch, ``Spatio-temporal deep learning-based undersampling artefact reduction for 2D radial cine MRI with limited training data,'' \emph{IEEE Trans. Med. Imaging}, vol. 39, no. 3, pp. 703--717, Aug. 2019.

\bibitem{b17} J. Schlemper, J. Caballero, J. V. Hajnal, A. N. Price, and D. Rueckert, ``A deep cascade of convolutional neural networks for dynamic MR image reconstruction,'' \emph{IEEE Trans. Med. Imaging}, vol. 37, no. 2, pp. 491--503, Oct. 2017.

\bibitem{b18} K. Hammernik \emph{et al., ``}Learning a variational network for reconstruction of accelerated MRI data,'' \emph{Magn. Reson. Med.}, vol. 79, no. 6, pp. 3055--3071, Jun. 2018.

\bibitem{b19} T. K{\"u}stner \emph{et al., ``}CINENet: deep learning-based 3D cardiac CINE MRI reconstruction with multi-coil complex-valued 4D spatio-temporal convolutions,'' \emph{Sci. Rep.}, vol. 10, no. 1, pp. 13710, Aug. 2020.

\bibitem{b20} T. Eo, Y. Jun, T. Kim, J. Jang, H. J. Lee, and D. Hwang, ``KIKI‐net: cross‐domain convolutional neural networks for reconstructing undersampled magnetic resonance images,'' \emph{Magn. Reson. Med.}, vol. 80, no. 5, pp. 2188--2201, Nov. 2018.

\bibitem{b21} H. El‐Rewaidy \emph{et al., ``}Multi‐domain convolutional neural network (MD‐CNN) for radial reconstruction of dynamic cardiac MRI,'' \emph{Magn. Reson. Med.}, vol. 85, no. 3, pp. 1195--1208, Mar. 2021.

\bibitem{b44} S. Wang, R. Wu, S. Jia, A. Diakite, C. Li, Q. Liu, H. Zheng, and L. Ying, ``Knowledge‐driven deep learning for fast MR imaging: Undersampled MR image reconstruction from supervised to un‐supervised learning.'' \emph{Magn. Reson. Med.}, vol. 92, no. 2, pp. 496--518, Aug. 2024.

\bibitem{b45} G. Zeng, Y. Guo, J. Zhan, Z. Wang, Z. Lai, X. Du, X. Qu, and D. Guo, ``A review on deep learning MRI reconstruction without fully sampled k-space.'' \emph{BMC Med. Imaging}, vol. 21, no. 1, pp. 195, Dec. 2021.

\bibitem{b46} R. Heckel, M. Jacob, A. Chaudhari, O. Perlman, and E. Shimron, ``Deep learning for accelerated and robust MRI reconstruction.'' \emph{Magn. Reson. Mater. Phy.}, pp. 1--34, Jul. 2024.

\bibitem{b47} V. Antun, F. Renna, C. Poon, B. Adcock, and A. C. Hansen, ``On instabilities of deep learning in image reconstruction and the potential costs of AI.'' \emph{Proc. Natl. Acad. Sci.}, vol. 117, no. 48, pp. 30088--30095, Dec. 2020.

\bibitem{b22} B. Yaman, S. A. H. Hosseini, S. Moeller, J. Ellermann, K. U{\u{g}}urbil, and M. Ak{\c{c}}akaya, ``Self‐supervised learning of physics‐guided reconstruction neural networks without fully sampled reference data,'' \emph{Magn. Reson. Med.}, vol. 84, no. 6, pp. 3172--3191, Dec. 2020.

\bibitem{b23} B. Yaman \emph{et al., ``}Multi‐mask self‐supervised learning for physics‐guided neural networks in highly accelerated magnetic resonance imaging,'' \emph{NMR Biomed.}, vol. 35, no. 12, pp. e4798, Dec. 2022.

\bibitem{b48} C. Millard, and M. Chiew, ``A theoretical framework for self-supervised MR image reconstruction using sub-sampling via variable density Noisier2Noise.'' \emph{IEEE Trans. Comput. Imaging}, vol. 9, Jul. 2023.

\bibitem{b49} N. Moran, D. Schmidt, Y. Zhong, and P. Coady, ``Noisier2noise: Learning to denoise from unpaired noisy data.'' \emph{IEEE/CVF Conf. Comput. Vis. Pattern Recognit.}, pp. 12064--12072, 2020.

\bibitem{b24} B. Zhou \emph{et al., ``}Dual-domain self-supervised learning for accelerated non-cartesian mri reconstruction,'' \emph{Med. Image Anal.}, vol. 81, pp. 102538, Oct. 2022.

\bibitem{b25} A. D. Desai \emph{et al., ``}Noise2Recon: Enabling SNR‐robust MRI reconstruction with semi‐supervised and self‐supervised learning,'' \emph{Magn. Reson. Med.}, vol. 90, no. 5, pp. 2052--2070, Nov. 2023.

\bibitem{b41} Y. Song, L. Shen, L. Xing, and S. Ermon, ``Solving inverse problems in medical imaging with score-based generative models,'' \emph{arXiv:2111.08005}, Nov. 2021.

\bibitem{b42} Z. X. Cui \emph{et al., ``}Self-score: self-supervised learning on score-based models for MRI reconstruction,'' \emph{arXiv:2209.00835}, Sep. 2022.

\bibitem{b43} A. Aali, M. Arvinte, S. Kumar, and J. I. Tamir, ``Solving inverse problems with score-based generative priors learned from noisy data,'' \emph{arXiv:2305.01166}, May. 2023.

\bibitem{b26} P. Bachman, R. D. Hjelm, and W. Buchwalter, ``Learning representations by maximizing mutual information across views,'' \emph{NeurIPS}, Vancouver, Canada, vol. 32, Dec. 2019.

\bibitem{b27} K. He, H. Fan, Y. Wu, S. Xie, and R. Girshick, ``Momentum contrast for unsupervised visual representation learning,'' \emph{Proc. CVPR}, Seattle, WA, USA, pp. 9729--9738, Jun. 2020.

\bibitem{b28} X. Chen, and K. He, ``Exploring simple siamese representation learning,'' \emph{Proc. CVPR}, pp. 15750--15758, Jun. 2021.

\bibitem{b29} A. Bardes, J. Ponce, and Y. LeCun, ``VICReg: variance-invariance-covariance regularization for self-supervised learning,'' \emph{ICLR}, Apr. 2022.

\bibitem{b30} J. Bromley, I. Guyon, Y. LeCun, E. S{\"a}ckinger, and R. Shah, ``Signature verification using a" siamese" time delay neural network,'' \emph{Adv. Neural Inf. Processing Syst.}, vol. 6, 1993.

\bibitem{b31} T. Chen, S. Kornblith, M. Norouzi, and G. Hinton, ``A simple framework for contrastive learning of visual representations,'' \emph{ICML}, Vienna, Austria, Nov. 2020.

\bibitem{b32} J. Zbontar, L. Jing, I. Misra, Y. LeCun, and S. Deny, ``Barlow twins: Self-supervised learning via redundancy reduction,'' \emph{ICML}, Jul. 2021.

\bibitem{reviewSeg} Z. Liu, K. Kainth, A. Zhou, T. W. Deyer,  Z. A. Fayad, H. Greenspan, and X. Mei, ``A review of self‐supervised, generative, and few‐shot deep learning methods for data‐limited magnetic resonance imaging segmentation,'' \emph{NMR Biomed.}, pp. e5143, Mar. 2024.

\bibitem{b33} S. Wang \emph{et al., ``}PARCEL: physics-based unsupervised contrastive representation learning for multi-coil MR imaging,'' \emph{IEEE/ACM Trans. Comput. Biol. Bioinform.}, Oct. 2022.

\bibitem{b34} Q. Yi, J. Liu, L. Hu, F. Fang, and G. Zhang, ``Contrastive learning for local and global learning mri reconstruction,'' \emph{arXiv:2111.15200}, Nov. 2021.

\bibitem{b35} A. V. D. Oord, Y. Li, and O. Vinyals, ``Representation learning with contrastive predictive coding,'' \emph{arXiv:1807.03748}, Jul. 2018.

\bibitem{b36} M. Arjovsky, A. Shah, and Y. Bengio, ``Unitary evolution recurrent neural networks,'' \emph{ICML}, New York City, NY, USA, Jun. 2016.

\bibitem{b37} R. Ahmad, H. Xue, S. Giri, Y. Ding, J. Craft, and O. P. Simonetti, ``Variable density incoherent spatiotemporal acquisition (VISTA) for highly accelerated cardiac MRI,'' \emph{Magn. Reson. Med.}, vol. 74, no. 5, pp. 1266--1278, Nov. 2015.

\bibitem{ocmr} C. Chen \emph{et al., ``}OCMR (v1. 0)--open-access multi-coil k-space dataset for cardiovascular magnetic resonance imaging.'' \emph{ArXiv Preprint ArXiv:2008.03410}, 2020.

\bibitem{b38} K. Hammernik, and T. K{\"u}stner, ``Machine enhanced reconstruction learning and interpretation networks (MERLIN),'' \emph{ISMRM}, London, UK, May. 2022.

\bibitem{b39} D. P. Kingma, and J. Ba, ``Adam: A method for stochastic optimization,'' \emph{arXiv:1412.6980}, Dec. 2014.

\bibitem{b40} S. G. Lingala, Y. Hu, E. DiBella, and M. Jacob, ``Accelerated dynamic MRI exploiting sparsity and low-rank structure: kt SLR,'' \emph{IEEE Trans. Med. Imaging}, vol. 30, no. 5, pp. 1042--1054, Jan. 2011.

\bibitem{bahadir2019learning} C. D. Bahadir, A. V. Dalca, and M. R. Sabuncu, ``Learning-based optimization of the under-sampling pattern in MRI,'' \emph{IPMI}, Proceedings 26, pp. 780--792, Jun. 2019.

\bibitem{zibetti2021fast} M. V. Zibetti, G. T. Herman, and R. R. Regatte, ``Fast data-driven learning of parallel MRI sampling patterns for large scale problems,'' \emph{Sci. Rep.}, vol. 11, no. 1, pp. 19312, Sep. 2021.

\end{thebibliography}
\end{document}